%% file: main.tex
\pdfoutput=1

\documentclass[9pt,journal,compsoc]{IEEEtran}

\usepackage{graphicx}
\usepackage{subfig}
\usepackage{comment}
\usepackage{multirow}
\usepackage{color}

\newcommand{\arch}{StochMem}

\newcommand{\ignore}[1]{}

\newcommand{\convlfsr}{{\em $Conv_{LFSR}$}}
\newcommand{\convmtj}{{\em $Conv_{MTJ}$}}
\newcommand{\convpwm}{{\em $Conv_{PWM}$}}
\newcommand{\pcmmtj}{{\em $PCM_{MTJ}$}}
\newcommand{\pcmpwm}{{\em $PCM_{PWM}$}}

\usepackage{color}
\newcommand{\redHL}[1]{\textcolor{red}{#1}}

\newcommand{\shrink}{Eliminate vertical white-space}
\newcommand{\vshrink}[1]{
  \ifdefined\shrink 
	\vspace{-#1cm}
  \else
	\vspace{0cm}
  \fi
}

\begin{document}


\ignore{
\title{Alternate {\ttlit ACM} SIG Proceedings Paper in LaTeX
Format\titlenote{(Produces the permission block, and
copyright information). For use with
SIG-ALTERNATE.CLS. Supported by ACM.}}
\subtitle{[Extended Abstract]
\titlenote{A full version of this paper is available as
\textit{Author's Guide to Preparing ACM SIG Proceedings Using
\LaTeX$2_\epsilon$\ and BibTeX} at
\texttt{www.acm.org/eaddress.htm}}}
}

\title{On Memory System Design for Stochastic Computing}

\author{S. Karen Khatamifard, M. Hassan Najafi, Ali Ghoreyshi, Ulya R. Karpuzcu, David Lilja}

\IEEEtitleabstractindextext{%
\begin{abstract}
\input{abs}

\end{abstract}


}

\maketitle


\ignore{
}

\section{Motivation}
\label{sec:intro}
\input{intro}


\section{Toward Seamless SC}
\label{sec:stochmem}

\input{stochmem}
\section{Evaluation Setup}
\label{sec:setup}
\input{setup}

\section{Evaluation}
\label{sec:eval}
\input{eval}

\section{Conclusion}
\label{sec:conc}
\input{conc}


\end{document}

%% file: abs.tex
Growing uncertainty in design parameters (and therefore, in design
functionality) 
renders
stochastic computing 
particularly promising, which represents and processes data as quantized
probabilities.  However, due to the difference in data representation,
integrating conventional memory (designed and optimized for non-stochastic
computing) in stochastic computing systems inevitably incurs a significant data
conversion overhead.  Barely any stochastic computing proposal to-date covers
the memory impact.  In this paper, as the first study of its kind to the best
of our knowledge, we rethink the memory system design for stochastic computing.  
%
The result is
a seamless stochastic system, \arch,
which features analog memory to trade
the energy and area overhead of data conversion 
for computation accuracy.  In this manner \arch\
can reduce the energy (area) overhead by up-to 52.8\% (93.7\%) at the cost of at
most 0.7\% loss in computation accuracy.

%% file: intro.tex
\vshrink{0.53}

\noindent
Stochastic Computing (SC)
has received renewed attention
in recent
years
~\cite{Alaghi_Survey,Armin_DAC2013,Hayes_DAC2015,Neural_Kim_DAC2016}. 
This is due to the growing {uncertainty} in design parameters, and therefore, in
design functionality,
as induced by 
imbalances in modern technology scaling. Representing and processing data as
{quantized probabilities}, SC becomes a natural fit. 
%
Data operands in SC 
take the form of 
bitstreams which encode probabilities:
independent of the length (and
interleaving of 0s and 1s), 
the ratio of the number of 1s to the length of the
bitstream determines the operand value.
%
Computation accuracy increases with the length of the bitstream at the cost of
higher-latency 
stochastic operations~\cite{Hayes_DAC2015}.
Still, computing with probabilities can reduce arithmetic complexity significantly,
such that
the hardware resource cost and the 
power consumption 
become orders of magnitude less than their conventional 
(i.e., non-stochastic) 
counterparts~\cite{Peng_TVLSI14,Najafi:2017:RAS:3098274.3060537}.
At the same time, computing with probabilities results in better tolerance to 
inaccuracy 
in input data operands~\cite{Hayes_DAC2015}.
The common focus of SC proposals from 1960s
onwards
has been stochastic logic (arithmetic), neglecting
memory, which 
represents a crucial system component.
%
Memory mainly serves as a repository for data collected from external resources (e.g.,
sensors) or data generated by previous steps of computation, to be used at later
stages of computation. 
Algorithmic characteristics 
dictate both, the memory capacity requirement and the memory access pattern
(particularly for data re-use).
%
Most SC proposals 
deploy conventional digital memories (designed and optimized for non-stochastic
computing) to
address such algorithmic needs.
Unfortunately, this practice increases hardware design complexity due to the 
discrepancy in conventional digital (i.e., non-stochastic) and stochastic data
representations. Digital to/from stochastic data conversion can
reach
80\% or more of the overall energy consumption and hardware cost, which can easily diminish any
benefit from stochastic computing~\cite{Armin_DAC2013,Weikang_2011}.  
{\em In this study, we rethink the memory system design  
for stochastic computing.}

\ignore{We will show that more than 80\%XXX 
of energy/area of a conventional stochastic system is wasted in these
conversions, diminishing the benefits of SC.}

Practically {\em seamless} conversion options between
analog and stochastic data representations~\cite{ASC1,ASC_MTJ_2015} makes analog
memory stand out as a particularly promising point in the memory design space for SC.  
%
%
The downside 
is potential loss in data accuracy,
where a divergence between the written/stored and the read
values (at the same memory address) often becomes inevitable, however, which stochastic
logic can mask due to its implicit tolerance to inaccuracy in input data
operands. 

\input{stochMemFig}

{\em This paper
quantitatively characterizes the potential of analog memory for seamless SC,
using a representative near-sensor stochastic image processing system as a case study\footnote{
\noindent Non-stochastic, analog near-sensor image processing accelerators such as~\cite{redeye} exist. The focus of
this paper is not design and exploration of image processing
accelerators. The scope rather is memory system design for stochastic computing where
we use a representative 
stochastic system to characterize the impact of memory.
}.
}
We will refer to the
resulting (practically) seamless stochastic system as \arch.
Cameras have already become ubiquitous sensors. There is a demand for near-sensor image processing
both to reduce costly communication with the cloud and to enhance security and
privacy.
Real-time image processing algorithms often track differences between a
stream of frames. It is not uncommon that the processing of the
instantaneous frame requires comparison to a history 
of previously processed frames, which has to be stored in and retrieved from some form of memory.
In the following, we will cover
five representative image processing applications which span diverse compute
and memory access characteristics.

%% file: stochMemFig.tex
\begin{figure*}[t]
  \centering
   \subfloat[Baseline Near-Sensor Stochastic Image Processor featuring Digital Memory] {
	\includegraphics[width=0.85\textwidth]{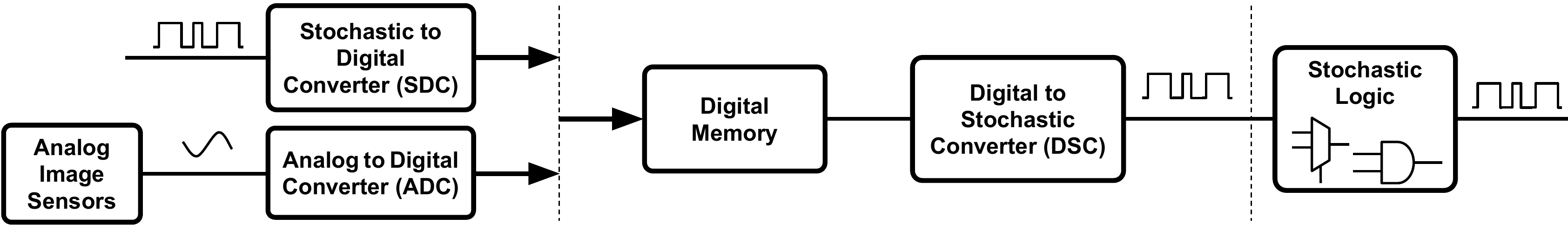}
	\label{fig:conv}
   }
   \\\vshrink{0.1}
   \subfloat[\arch\ featuring Analog Memory] {
	\includegraphics[width=0.85\textwidth]{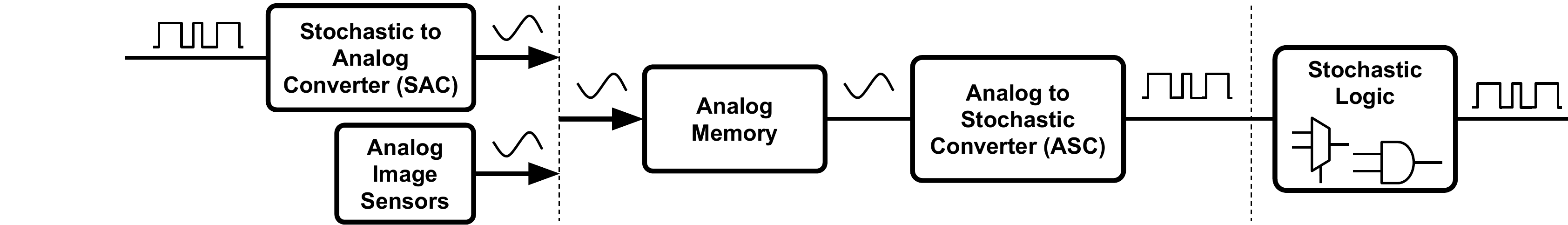}
	\label{fig:analog}
   }
   \vshrink{0.2}
	\caption{Baseline Near-Sensor Stochastic Image Processor vs. \arch.}
	\label{fig:StochMemDiagram}
   \vshrink{0.5}
\end{figure*}

%% file: stochmem.tex
\noindent We will first compare and contrast \arch\ featuring analog memory 
with 
the corresponding stochastic near-sensor image processor 
featuring conventional digital memory 
as a representative baseline. 
\vshrink{0.3}



\subsection{Baseline: Stochastic Logic + Conventional Memory}
\label{sec:convstoch}
\noindent 
Fig.~\ref{fig:conv} provides an overview for a 
the baseline 
stochastic near-sensor image processor
featuring conventional digital memory (designed and optimized for non-stochastic
computing).
The input data operands may represent the result bitstreams of previous steps of (stochastic)
computation, or may directly come from analog image sensors.  
To be able to store such input data in conventional digital memory,  
a {\em Stochastic to Digital Converter}, SDC (for stochastic input bitstreams) or an {\em Analog to Digital
Converter}, ADC (for analog inputs coming from sensors) become necessary.
Moreover, further (stochastic) processing of the stored data
necessitates a {\em 
Digital to Stochastic Converter}, DSC, upon data retrieval from digital memory.
In the following we briefly describe key system components.

\noindent \textbf{Stochastic Logic} 
incorporates a 
circuit
of 
basic Boolean gates 
to carry out the 
application-specific  stochastic computation (Section~\ref{sec:bac}).
The inputs and outputs are both stochastic bitstreams.

\noindent \textbf{Stochastic to Digital Converter (SDC)} can generate the
conventional binary representation for any stochastic bitstream. A digital counter usually
serves the purpose, by keeping track of the number of 1s in the input bitstream
to be converted.
An SDC carries out data conversion if the inputs to the stochastic system represent result bitstreams from
previous steps of (stochastic) computation.

\noindent \textbf{Analog to Digital Converter (ADC)} 
becomes necessary if the inputs to the stochastic system directly come from analog
image sensors.
Conventional ADCs
can serve the purpose.
For most applications of SC (including the case study in this paper) an 8 to 10-bit ADC is
sufficient~\cite{Peng_TVLSI14}.   
  
\noindent \textbf{Digital to Stochastic Converter (DSC)}
transforms conventional binary data retrieved from digital memory (for further stochastic
processing) to stochastic bitstreams. Commonly, DSC achieves this by 
comparing an unbiased random number (obtained from a random number generator) to
the binary value to be converted. A one is attached to the output
(stochastic) bitstream if the random number is
less than the binary value (to be converted); zero, otherwise.
The random number generator can rely on physical random sources
or
pseudo-random constructs such as Linear Feedback Shift Registers (LFSRs).
\vshrink{0.4}
\ignore{
Another approach to convert a digital input into a stochastic signals is 
to first use a Digital-to-Analog converter (DAC) and produce an analog equivalent of the input value, and then
use an Analog-to-Stochastic Converter (ASC) to generate the corresponding
stochastic signal. We discuss ASCs in the next section. 
}

\subsection{StochMem: Stochastic Logic + Analog Memory}
\label{sec:stochmem2}
\noindent 
The data converters (SDC or ADC and DSC) incorporated
into the baseline stochastic system from Fig.~\ref{fig:conv}
each has a significant energy
and area 
footprint~\cite{Armin_DAC2013}, 
which can easily nullify potential benefits from SC.
In order to reduce this overhead, \arch\ replaces the conventional digital
memory with its analog counterpart.
Fig.~\ref{fig:analog} provides the overview for the resulting SC system.
\ignore{used in proposed system to
transform the data between the analog and the stochastic representations:}
In the following we briefly describe key \arch\ components:

\noindent \textbf{Stochastic Logic} is the same as under the baseline 
system.

\noindent \textbf{Stochastic to Analog Converter (SAC)} replaces the SDC of the
conventional system.  SAC can generate the analog representation for any stochastic
bitstream.  A conventional 
analog integrator can serve the purpose,
by measuring the
fraction of time 
a stochastic input bitstream stays at logic 1. 
Such an integrator usually has a smaller energy and area footprint than the SDC of the
baseline
system (Section~\ref{sec:setupHW}). A SAC carries
out data conversion if the inputs to \arch\ represent result
bitstreams from previous steps of (stochastic) computation.

\noindent \textbf{Analog to Stochastic Converter (ASC)} 
transforms 
data 
from analog memory (for further stochastic
processing) to stochastic bitstreams, similar to the DSC of the conventional
system.
As representative examples,~\cite{ASC1, ASC_MTJ_2015} both cover energy-efficient ways
for generating 
stochastic bitstreams 
from analog inputs.

%% file: setup.tex
\subsection{System Design}
\label{sec:sd}

\noindent We evaluate three stochastic near-sensor image-processing designs: two different
implementations of the baseline from Fig.~\ref{fig:conv} (\convlfsr\
and \convmtj, respectively) and \arch. The two baseline designs differ in the
implementation of data converters as follows: 

\noindent {\convlfsr:} 
The baseline SC system featuring 
a 
10-bit LFSR and a comparator
as the DSC unit.

\noindent {\convmtj:} 
The baseline featuring  
a DAC followed by an MTJ-based ASC
as a more energy-efficient DSC. The rest of
the system is identical to \convlfsr.
\vshrink{0.3}

\ignore{
\noindent {\convpwm:}
The baseline system featuring 
a DAC followed by a PWM-based ASC 
as the DSC unit.
}
\ignore{
\noindent {\em \arch:}
\arch\ featuring an analog memory 
 and an MTJ-based ASC.
 \vshrink{0.4}
}
\ignore{
\noindent {\pcmpwm:}
\arch\ featuring  an analog PCM (Section~\ref{sec:pcm}) and
a PWM-based ASC.
}

\ignore{
We use stochastic logic implementations of five image processing applications.
{\em Robert} (Robert's cross edge detection),
{\em Median} (median filter noise reduction),
and {\em Frame}
(frame difference-based image segmentation) from~\cite{Peng_TVLSI14};
{\em Gamma} (gamma correction) from~\cite{Weikang_2011}; 
{\em KDE} (kernel density estimation-based image 
segmentation) from~\cite{Peng_TVLSI14}.
\footnote{
Due to page limitations, we do not show the circuits.}
} 


\subsection{Stochastic Applications}
\label{sec:bac}

\noindent 
To evaluate {\em Stochastic Logic} from Fig.~\ref{fig:StochMemDiagram}, we use
stochastic circuits of five representative image processing applications:
{\em Robert} (Robert's cross edge detection),
{\em Median} (median filter noise reduction),
{\em Frame}
(frame difference-based image segmentation) from~\cite{Peng_TVLSI14};
{\em Gamma} (gamma correction) from~\cite{Weikang_2011}; and
{\em KDE} (kernel density estimation-based image 
segmentation) from~\cite{Peng_KDE_2011}.

\ignore{
We use stochastic implementations of five image processing applications.
{\em Robert}, the Robert's cross edge detection circuit
from~\cite{Peng_TVLSI14}; 
{\em Median}, the median filter noise reduction circuit
from~\cite{Peng_TVLSI14};
{\em Frame},
the frame difference-based image segmentation circuit from~\cite{Peng_TVLSI14};
{\em Gamma}, the gamma correction circuit from~\cite{Weikang_2011}; and 
{\em KDE}, the kernel density estimation-based (KDE) image 
segmentation circuit from~\cite{Peng_KDE_2011}.
}

\ignore{
We use stochastic implementations of five image processing applications:
{\em Robert}, the Robert's cross edge detection; 
{\em Median}, the median filter noise reduction;
{\em Frame},
the frame difference-based image segmentation;
{\em Gamma}, the gamma correction; and 
{\em KDE}, the kernel density estimation-based (KDE) image 
segmentation.
}



\vspace{-1em}

\begin{figure}[h!]
	\centering
		\hspace{-0.2cm}
		\includegraphics[height=1.43cm]{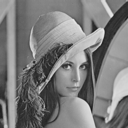}
		\hspace{0.01cm}
		\includegraphics[height=1.43cm]{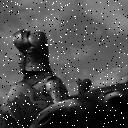} 
		\hspace{0.01cm}
		\includegraphics[height=1.43cm]{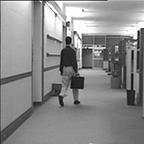} 
		\hspace{0.01cm}
		\includegraphics[height=1.43cm]{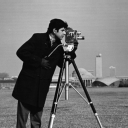} 
		\hspace{0.01cm}
		\includegraphics[height=1.43cm]{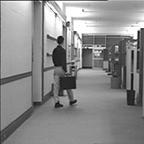} 

\vshrink{0.3}

	\subfloat[{\em Robert}] {
		\includegraphics[height=1.43cm]{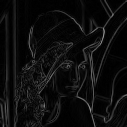}
	}   
	\subfloat[{\em Median}] {
		\includegraphics[height=1.43cm]{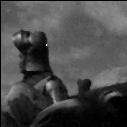} 
	}   
	\subfloat[{\em Frame}] {
		\includegraphics[height=1.43cm]{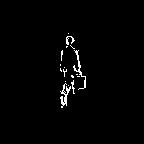} 
	}   
	\subfloat[{\em Gamma}] {
		\includegraphics[height=1.43cm]{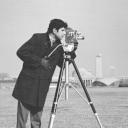} 
	}   
	\subfloat[{\em KDE}] {
		\includegraphics[height=1.43cm]{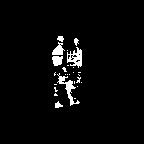} 
	}   
	\vshrink{0.1}
	\caption{Input (expected output) 
	per application on top (bottom).}
	\label{fig:input}
	\vshrink{.2}
\end{figure}
\input{setuptbl-date17}


As input, we use $128\times128$ gray-scale images for 
{\em Robert, Median, Frame}, and {\em Gamma}; and 33 recent frames of a video, for {\em KDE}. 
Fig.~\ref{fig:input} shows the input (expected output) images used for each
application on
the top
(bottom) row. Expected output captures the maximum-possible accuracy.
To calculate the accuracy of the end results, 
we calculate the average pixel-by-pixel difference
between the output image of each stochastic circuit and the corresponding 
maximum-possible-accuracy output. 
\vshrink{.4}

\subsection{Hardware Parameters}
\label{sec:setupHW}


\noindent 
Table~\ref{tbl:setup} summarizes the 
area and energy consumption of different units of the
evaluated stochastic systems. 
We synthesize logic units (including the stochastic circuit implementations 
of the five benchmark applications from Section~\ref{sec:bac}), LFSR,
digital comparator, and counter units using Synopsys Design
Compiler vH2013.12 with a 45nm gate library.
The analog memory implementation follows~\cite{analogmem}.
\input{accur_fig}

For a fair evaluation, we assume that the input to both the baseline
designs and \arch\
directly comes from analog image sensors. 
All designs output a stochastic bitstream.
Therefore, the evaluated systems do not feature an SDC or SAC on the feedback
path from memory (Fig.~\ref{fig:StochMemDiagram}). However, we include these units 
in Table~\ref{tbl:setup} for the sake of completeness.
SAC area (energy) cost
is 2.3$\times$ (17.9$\times$) less than SDC. 
Accordingly, if the evaluated systems deployed these units (as explained in Section~\ref{sec:stochmem}), 
\arch\ would have shown 
even larger gains when compared to the baseline.
\vshrink{0.3}

\ignore{
\noindent{\bf Sensitivity evaluation:}

For each sample image, for each inaccuracy rate, we generate 20 degraded
versions of each sample image. Input pixel intensities of each degraded input image are
scaled down to [0,1] interval and then
converted to stochastic streams of 128 to 1024 random bits. We process the stochastic bit-stream 
corresponding to each image pixel intensity by feeding it to a stochastic circuit. 
The average output error rate for the output image produced by each circuit is computed as follows:
$$E = \frac{\sum_{i=1}^{W} \sum_{j=1}^{H} |T_{i,j}-S_{i,j}|}{(W\times H) } \times 100$$
where $S_{i,j}$ is the expected pixel value in the golden output image (the output of processing the original
sample input) and $T_{i,j}$ is the pixel value produced using the circuit when processing a degraded image.
}

%% file: setuptbl-date17.tex
\begin{table}[h]
\centering
\caption{Area and energy breakdown.}
\label{tbl:setup}
\resizebox{0.86\columnwidth}{!}{
\begin{tabular}{ccc}
\hline\hline
\multicolumn{3}{c}{Stochastic Logic} \\
\hline
Circuit & Area ($um^2$) & Energy (pJ)(@1GHz)\\
\hline
Robert & 339 & 0.440 \\
Median & 5382 & 4.090 \\
Frame & 457 & 0.413 \\
Gamma & 76 & 0.042 \\
KDE & 8691 & 7.094 \\
\hline\hline
\multicolumn{3}{c}{Baseline System Parameters} \\
\hline
Unit & Area ($um^2$) & Energy (pJ)(@1GHz)\\
\hline
ADC 10-bit~\cite{ADCSurvey} & 50,000 & 20 \\
SRAM cell & 0.35 & 10 \\
DSC: 10-bit LFSR & 194 & 0.355 \\
DSC: 10-bit Comparator & 96 & 0.041 \\
DSC: DAC 8-bit~\cite{DACSurvey} & 16,000 & 64 \\
SDC: 10-bit Counter & 254 & 0.179 \\
\hline\hline
\multicolumn{3}{c}{StochMem System Parameters} \\
\hline
Unit & Area ($um^2$) & Energy (pJ)(@1GHz)\\
\hline
Analog memory cell~\cite{analogmem} & 58.7 & 10 (RD) / 100 (WR) \\
ASC ~\cite{ASC_MTJ_2015} & 15 & 0.030 \\
SAC (integrator) & 110 & 0.010 \\
\hline\hline
\end{tabular}
}
\end{table}

%% file: accur_fig.tex
\begin{figure*}[ht]
	\centering
  \begin{minipage}{0.46\textwidth}
	\centering
	\includegraphics[width=0.8\columnwidth]{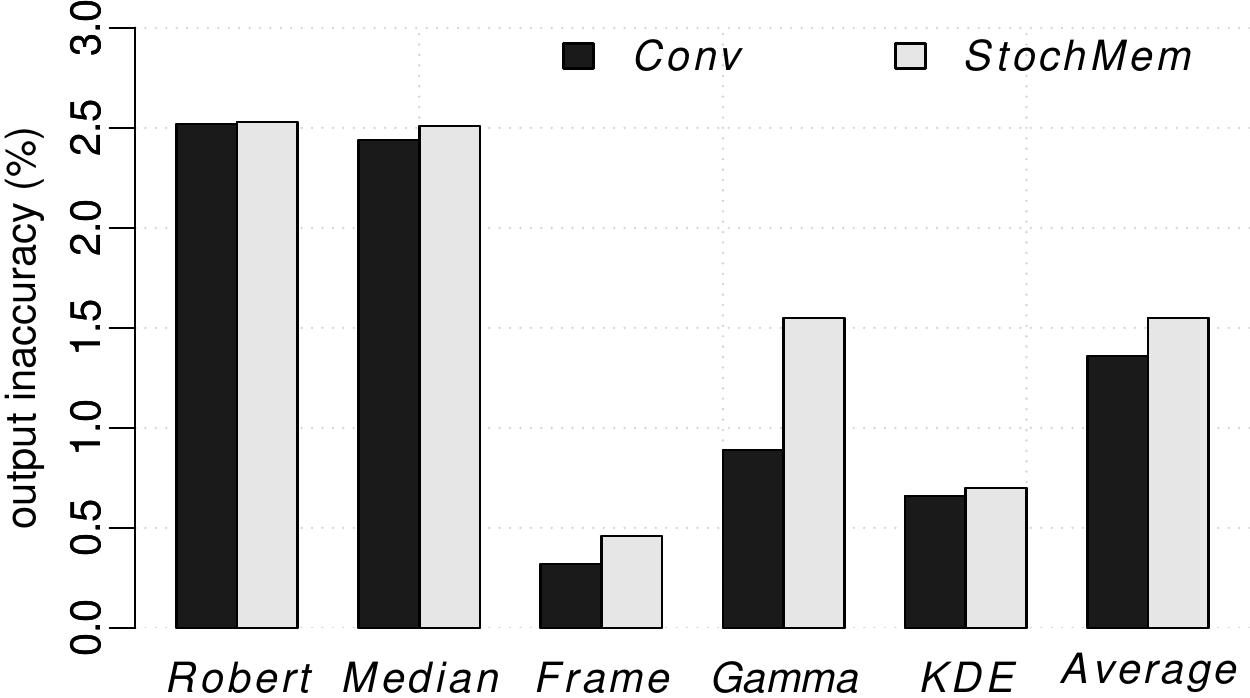}
	\caption{Output inaccuracy of the baseline vs. \arch.}
	\label{fig:accur}
  \end{minipage}
  \begin{minipage}{0.53\textwidth}
    \centering	
	\subfloat[{\em Robert}] {
		\includegraphics[height=3.5cm]{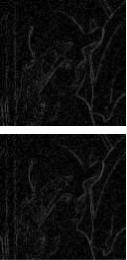}
	}   
	\hspace{-0.3cm}
	\subfloat[{\em Median}] {
		\includegraphics[height=3.5cm]{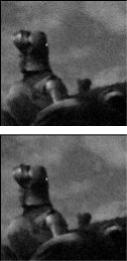} 
	}
	\hspace{-0.3cm}   
	\subfloat[{\em Frame}] {
		\includegraphics[height=3.5cm]{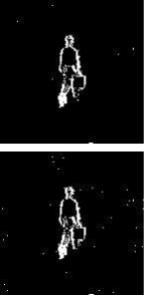} 
	}
	\hspace{-0.3cm}   
	\subfloat[{\em Gamma}] {
		\includegraphics[height=3.5cm]{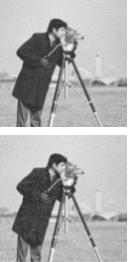} 
	}
	\hspace{-0.3cm}   
	\subfloat[{\em KDE}] {
		\includegraphics[height=3.5cm]{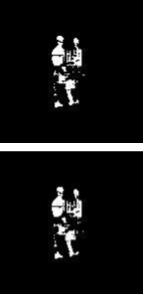} 
	}   
	\vshrink{0.1}
	\caption{Output images: Baseline (\arch) on top (bottom).}
	\label{fig:outputs}
  \end{minipage}
\end{figure*}

%% file: eval.tex
\noindent We start the evaluation with a quantitative characterization of the
accuracy loss in the outputs due to the
potential read-write discrepancy of the
analog
memory incorporated 
in \arch.
We continue with energy consumption and
conclude 
with area cost.
\vshrink{0.5}



\subsection{Output Accuracy of \arch}
\input{accur}
\label{sec:evalaccur}

\subsection{Reduction in Energy Consumption}
\input{energy}
\label{sec:energy}

\subsection{Reduction in Area}
\input{area}

\ignore{
\begin{figure*}[]
	\centering
	\includegraphics[height=2.7cm]{figs/AREAENERGY.jpg}
	\label{fig:convenergy}
	\caption{Area cost comparison among different applications and system designs.}
	\label{fig:area}
\end{figure*}
}

%% file: accur.tex
\noindent 
A known downside of analog memory technologies is the potential discrepancy 
between values read and written/stored. 
We model the impact of this discrepancy after the accuracy measurements of a
representative analog memory implementation~\cite{analogmem}.
All evaluated benchmark applications 
produce images as output. Therefore, 
we capture the accuracy loss in the output by the average per-pixel deviation
from the ``{expected}'' output, which corresponds to the maximum-possible
achievable accuracy for each application as shown in
the bottom row of Fig.~\ref{fig:input}.


Fig.~\ref{fig:accur} demonstrates the \% output inaccuracy of \arch\  and
the baseline designs for all applications under a stochastic bitstream length of
1024. The y-axis is normalized to the expected (i.e., maximum-possible achievable) accuracy values
corresponding to the images in the bottom row of  Fig.~\ref{fig:input}.
The two baseline designs evaluated, \convlfsr\ and \convmtj\
(Section~\ref{sec:sd}), feature
the very same output inaccuracy, as given by the {\em Conv} bar in
Fig.~\ref{fig:accur}.
We observe that, overall, 
the degradation (with respect to {\em Conv}) in the output accuracy of \arch\ remains
negligible.
Only for {\em Gamma}, the inaccuracy becomes around 0.7\% worse than {\em Conv}. For all other
applications, the inaccuracy worsens by less than 0.15\%. On average, the \%
output inaccuracy of \arch\ is 1.55\%; of {\em Conv}, 1.36\%, with respect to the
maximum-possible achievable output accuracy.
\ignore{
with respect to 
Besides, on average
output inaccuracy under \arch is a little worse than conventional, 1.55\% compared
to 1.36\%. 
}
Fig.~\ref{fig:outputs} tabulates the output images for all benchmark applications
under \arch\ and {\em Conv}. In accordance with the comparison results from
Fig.~\ref{fig:accur}, the difference in output accuracy is barely perceivable.
\ignore{
We repeat each experiment 20 times for statistical significance, and report the
median 
Since experiments are repeated 20 times,
after sorting the output of the applications based on measured
inaccuracy, we pick and show the median outputs in Fig.~\ref{fig:outputs}.
As depicted,
}

We repeat these experiments for 
3 different bitstream lengths: 128, 256, and 512 bits.
The average output inaccuracy of \arch\ with respect to {\em Conv} increases 
from 4.08\% to 4.21\%, from 2.63\% to 2.77\%, and from 1.87\% to 2.03\%,
 as the bitstream length increases 
from 128 to 512, respectively.
The relatively small degradation in the output inaccuracy is in line with the
experimental outcomes summarized in Fig.s~\ref{fig:accur}
and~\ref{fig:outputs}.
\vshrink{.5}

%% file: energy.tex
\begin{figure}[]
	\centering
	\includegraphics[width=0.9\columnwidth]{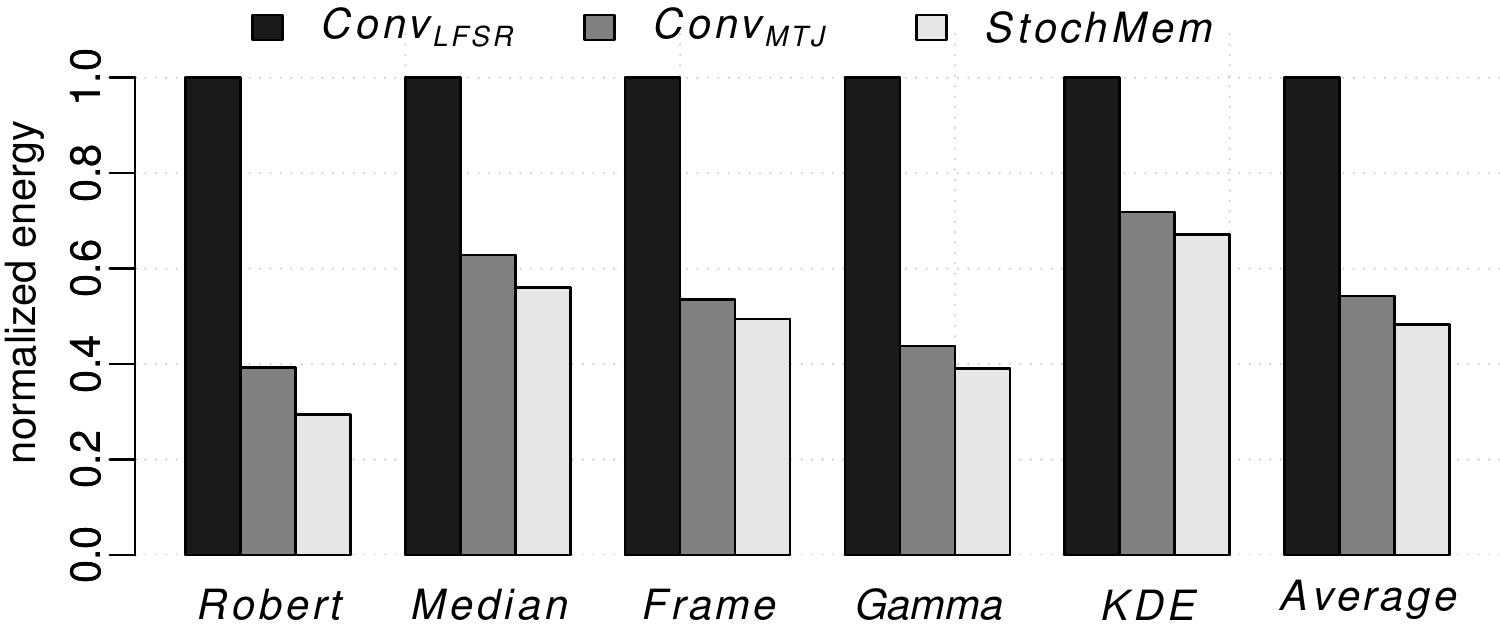}
	\caption{Energy consumption 
	normalized to \convlfsr.}
	\label{fig:energy}
	\vshrink{0.2}
\end{figure}

\noindent We next compare and contrast the energy consumption of 
the evaluated stochastic designs. 
In the following, we report the experimental results for a 
bitstream length of 1024 without loss of generality.
As Fig.~\ref{fig:energy} depicts, due to its more energy-efficient DSC
implementation, \convmtj\
can decrease 
the energy consumption 
with respect to \convlfsr\ significantly,
by 45.7\% on average. 
Introducing analog memory  -- i.e., \arch\ -- can reduce the energy consumption
further, by 11.1\% on average over \convmtj.
\ignore{
using DAC followed by an ASC (\convmtj) can decrease
the energy consumption of 
a conventional SC system (over \convlfsr) significantly,
by 51.3\% on average.
Introducing analog PCM memory (\pcmmtj) can improve the energy efficiency
further, by 3.6\% over \convmtj.
}

\ignore{ Since the PWM-based methods run for a much shorter time\footnote{The
accuracy of a PWM-based system running for 5ns is equivalent to an LFSR-based
system running for 1024ns\cite{PWM-ASPDAC2017}.}, we evaluate them separately in
Figure~\ref{fig:energypwm}, while rest of the methods are compared in
Figure~\ref{fig:energyconv}.  In each figure, y-axis represents normalized
energy consumption (to the most energy consuming conventional method). Different
applications are shown on the x-axis. Each bar in the figures represents one of
the five alternative systems.  As Figure~\ref{fig:energyconv} depicts, although
using DAC followed by an ASC (\convmtj) decreases the energy consumption of
system compared to \convlfsr significantly (by 51.3\% on average), using PCM
memory in \pcmmtj improves the energy efficiency further by 3.6\% over \convmtj.
A similar trend is observed in Table~\ref{tbl:energy}, where \pcmpwm is on
average 48.1\% more energy efficient than \convpwm.  
}

To demonstrate the sources of these energy gains, we quantify the share of
energy spent in different units.
We expect an energy-efficient 
stochastic system
to spend most of its energy budget on
computation, rather than on data conversion and input operand retrieval.
Pie charts from Fig.~\ref{fig:pie} differentiate between the shares of energy
spent in the  
{\em input layer} (which covers the input operand retrieval and hence
constitutes the ADC, if applicable, and memory units); 
in the {\em conversion units} (which constitute the ASC or DSC);
and in the {\em stochastic logic} (which captures the actual computation).
Figures~\ref{fig:pie1},~\ref{fig:pie2}, and \ref{fig:pie4}, show the shares
for {\convlfsr}, \convmtj, and \arch\ separately (Section~\ref{sec:sd}). 
As the charts reveal,
share of {\em stochastic logic} ({\em conversion units}) increases
(decreases) from 31.2\% (64.4\%) to 53.0\% (37.8\%) and to 60.1\% (22.1\%), as
we move from
\convlfsr\ to \convmtj\ and to \arch\ respectively. 
\arch\ represents the most
energy efficient design, featuring the lowest (highest) energy share for
data conversion (computation), when compared
to \convlfsr\ and \convmtj.
\vshrink{0.4}

\input{energyfigs}

\ignore{
%
Energy consumption analysis clearly shows
how using analog
memory in SC systems can minimize the energy overhead of conversion to improve
the overall energy efficiency.
\vshrink{0.44}
}

%% file: energyfigs.tex


\begin{figure}[tp]
	\centering
   \includegraphics[width=0.8\columnwidth]{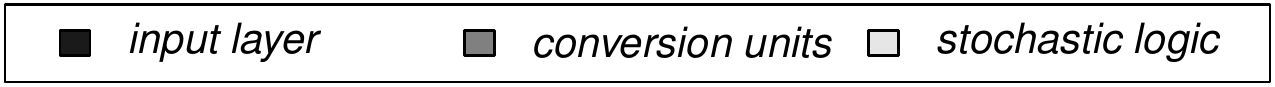} \vspace{-0.3cm} \\
   \vspace{-0.2cm}
   \subfloat[\convlfsr] {
	\includegraphics[width=0.32\columnwidth]{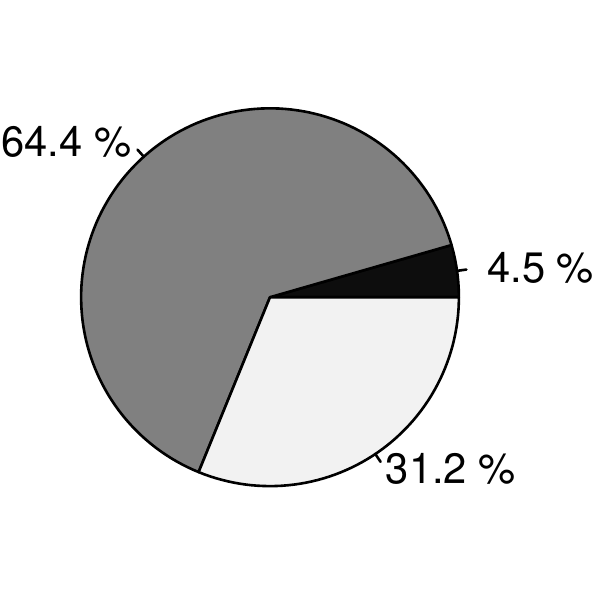}  
	\label{fig:pie1}
   }   
   \subfloat[\convmtj] {
   \hspace{-0.3cm}
	\includegraphics[width=0.32\columnwidth]{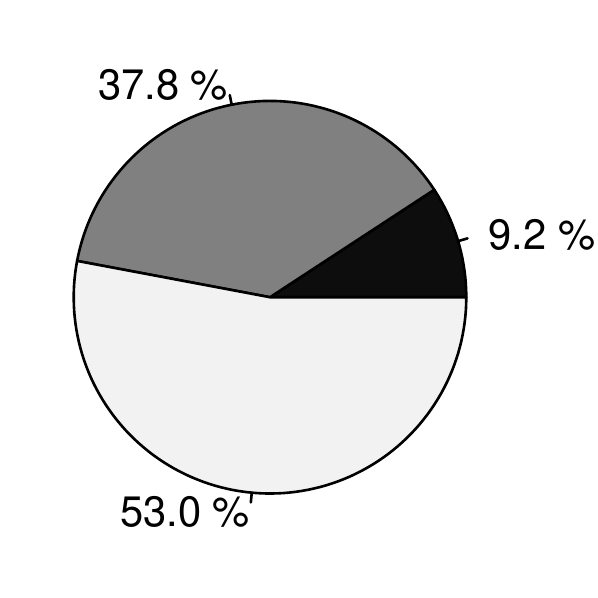}  
	\label{fig:pie2}
   }
   \subfloat[\arch] {
   \hspace{-0.3cm}
	\includegraphics[width=0.32\columnwidth]{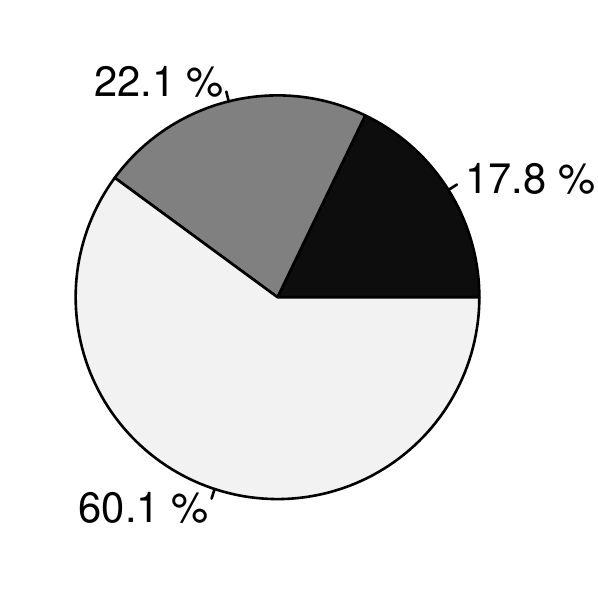} 
	\label{fig:pie4}
   }
	\caption{Share of energy consumed by different units.}
	\label{fig:pie}
 	\vshrink{0.5}
\end{figure}

%% file: area.tex
\input{areatbl}

\noindent In this section, we evaluate the area cost of each alternative. Since
tailoring ADC and DAC units to each application was out of the scope of this
study,
for the 
baseline (i.e., \convlfsr\ and \convmtj)
we 
deploy an ADC and a DAC unit of minimal area 
(which represents the
hypothetical best-case in terms of area cost),
even if
these units fail short of providing the required precision.
Accordingly, if we were to incorporate realistic ADC or DAC units 
(which would likely incur a much higher area overhead),
\arch\ (which does not employ any ADC or DAC) would have shown even
larger area savings in comparison to the baseline. 

\input{areafigs}

Table~\ref{tbl:area} summarizes the area cost for the evaluated stochastic
designs 
(columns) for the stochastic benchmark applications (rows). 
While \convmtj\ consumes notably less energy than  \convlfsr\
(Section~\ref{sec:energy}), it requires an extra DAC which increases the area
overhead (with respect to \convlfsr) by 20.0\% on average.
\ignore{
\convmtj\ can reduce
the energy consumption by 51.3\% over \convlfsr\ (Table~\ref{tbl:energy}),
however \convmtj\ needs an extra DAC unit which incurs
a higher area cost,
on average 20.0\% more. }
On the other hand, 
\arch\ can cut the area cost
significantly, by about 93.7\% (with respect to \convlfsr) on average, by
eliminating the need for costly conversion units.
Only \arch\ can deliver area and energy benefits at the same time. 

Fig.~\ref{fig:piearea} depicts a detailed break-down of
area consumption among different units.
Similar to Fig.~\ref{fig:pie},
pie charts from Fig.~\ref{fig:piearea} differentiate between the shares of 
area in the  
{\em input layer},
{\em conversion units},
and {\em stochastic logic}, respectively.
Only 4.9\% of the area in \convlfsr\
goes to the {\em stochastic logic}, while the {\em input layer} consumes 90.9\%.
{\em Stochastic logic} in \convmtj\ has even a smaller share of area
(4.1\%) when compared to \convlfsr.
On the other hand, in \arch,  63.1\% of the area goes
to {\em stochastic logic}; only 10.8\%, to {\em conversion units}.


Data conversion in conventional
SC systems necessitates high-overhead units such as LFSRs+comparators, ADCs, or DACs. 
\arch-like SC systems, on the other hand, 
can eliminate or replace these units with
lighter-weight counterparts 
leading to substantial energy
and area savings.
\vshrink{0.3}

%% file: areatbl.tex
\begin{table*}[htp]

	\centering
	\resizebox{.7\textwidth}{!}{
	\hspace{-0.5cm}
		\setlength{\tabcolsep}{0.3em} 
		{\renewcommand{\arraystretch}{1.2}
			\begin{tabular}{|c||c||c|c|c|c||c|c|c|c|c||c|c|c|}
				\hline
				\multirow{2}{*}{Apps} &
			  \multirow{2}{*}{\begin{tabular}[c]{@{}c@{}}Logic\end{tabular}} &
				\multicolumn{4}{c||}{\boldmath{\convlfsr}} &
				\multicolumn{5}{c||}{\boldmath{\convmtj}} &
				\multicolumn{3}{c|}{\bf{\em StochMem}} \\ 
				\cline{3-14} 
				&  & Memory & ADC & DSC & Total & Memory & ADC & DAC & ASC & Total & Memory & ASC & Total \\ \hline
				
				{\em Robert} & 339 & 21 &\multirow{5}{*}{50000} & 1450 & 51810 & 21 & \multirow{5}{*}{50000} & \multirow{5}{*}{16000} & 75 & 66435 & 183 & 75 & 597 \\ 
				\cline{1-3} \cline{5-7} \cline{10-14} 
				{\em Median} & 5382 & 38 & & 2900 & 58320 & 38 &  &  & 150 & 71570 & 336 & 150 & 5868 \\ 
				\cline{1-3} \cline{5-7} \cline{10-14}
				{\em Frame} & 457 & 17 &  & 772 & 51246 & 17 &  &  & 60 & 66534 & 153 & 60 & 670 \\ 
				\cline{1-3} \cline{5-7} \cline{10-14}
				{\em Gamma} & 76 & 35 &  & 1156 & 51267 & 35 &  &  & 120 & 66231 & 306 & 120 & 502 \\ 
				\cline{1-3} \cline{5-7} \cline{10-14}
				{\em KDE} & 8691 & 122 &  & 6166 & 64979 & 122 &  &  & 630 & 75443 & 1071 & 630 & 10392 \\ 
				\hline
				
			\end{tabular}
		}
	}
	\caption{Area in $\mu m^2$.\label{tbl:area}}
\vshrink{.5}
\end{table*}

%% file: areafigs.tex


\begin{figure}[tp]
	\centering
   \includegraphics[width=0.8\columnwidth]{./legend.pdf} \vspace{-0.3cm} \\
   \vspace{-0.2cm}
   \subfloat[\convlfsr] {
	\includegraphics[width=0.32\columnwidth]{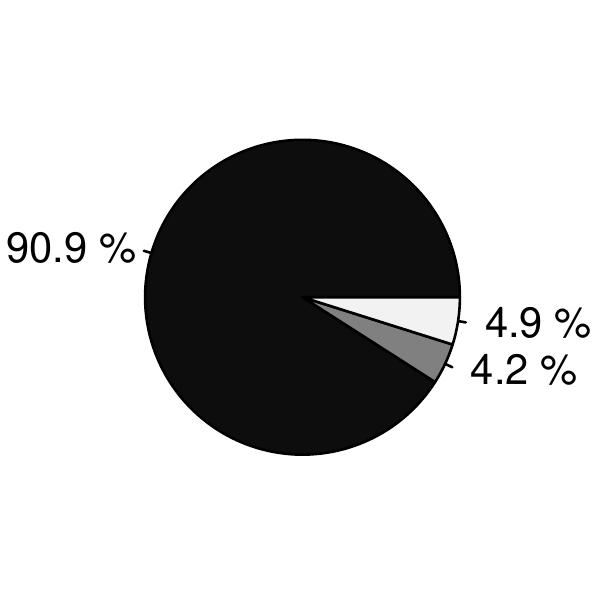}  
	\label{fig:piearea1}
   }   
   \subfloat[\convmtj] {
   \hspace{-0.3cm}
	\includegraphics[width=0.32\columnwidth]{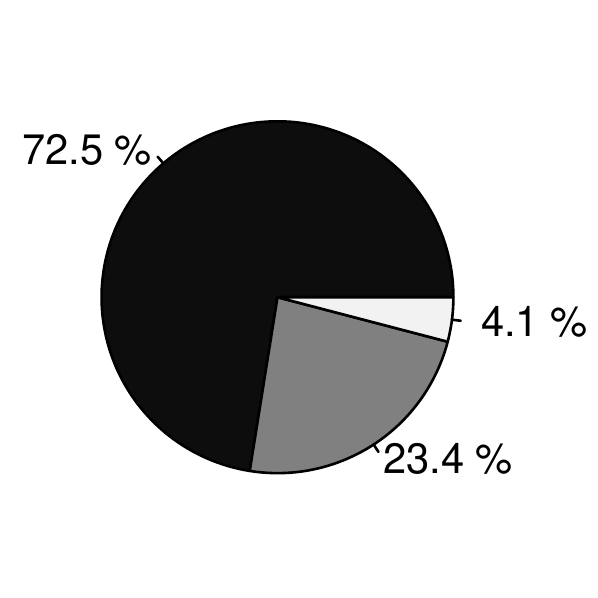}  
	\label{fig:piearea2}
   }
   \subfloat[\arch] {
   \hspace{-0.3cm}
	\includegraphics[width=0.32\columnwidth]{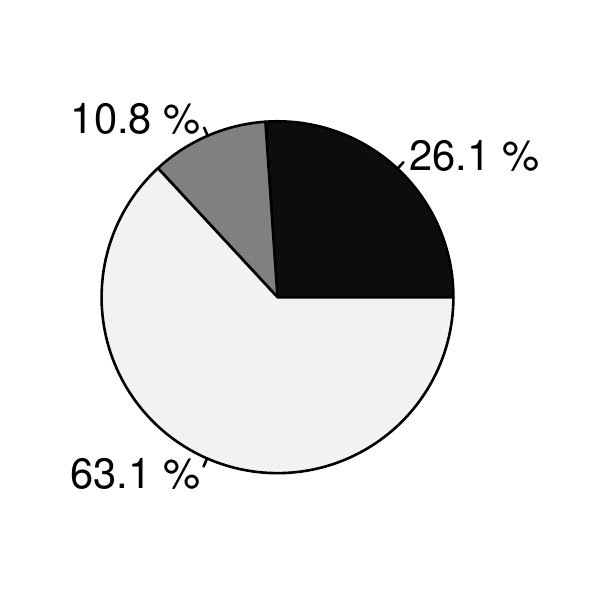} 
	\label{fig:piearea3}
   }
	\caption{Pie-charts demonstrating share of hardware cost (in terms of area) across different units.}
	\label{fig:piearea}
 	\vshrink{0.5}
\end{figure}

%% file: conc.tex
\noindent
A challenging artifact of modern technology scaling is growing uncertainty in
design parameters, and therefore, in design functionality. This renders
stochastic computing (SC) a particularly promising paradigm, which represents and processes
information
as quantized probabilities. Numerous stochastic computing proposals from 1960s
onwards, however, focus on stochastic logic (mainly arithmetic), neglecting
memory. Unfortunately, deploying conventional (digital) memory in a stochastic system
is particularly
inefficient due to the difference in data representations, which can easily
incur a significant data conversion overhead.

In this study, we rethink the memory system design for stochastic computing to minimize the
data conversion overhead, which can reach 80\% of overall hardware cost.  
Analog memory is particularly promising due to seamless conversion options
between analog and stochastic data representations, despite the potential loss
in data accuracy which stochastic logic can easily mask due to its implicit
fault tolerance.
We thus evaluate analog memory for
seamless SC, using a representative stochastic near-sensor image processing
system as a case study.
%
%
%
We demonstrate how
such a system 
%
can reduce energy consumption and area cost by up to 52.8\% and
93.7\%,
while keeping the accuracy loss as incurred by analog memory below 0.7\%.

\ignore{
Stochastic computing (SC) is a promising paradigm since it can
naturally address the growing
uncertainty in design parameters, and hence, functionality, a crucial artifact
of modern technology scaling.  In stochastic
computing, data is represented by quantized probabilities, rather than common
binary representation.  Conventionally, stochastic systems employ classic
memory. Hence, conversion from/to binary representation is needed in such
systems, consuming easily more than 80\% of overall hardware cost. Hence, we
propose StochMem, stochastic system with analog memory, to reduce conversion
overheads, leading to a more efficient stochastic system, with the cost of
degrading system's accuracy.
}